\documentclass[aps,prl,10pt,twocolumn,nofootinbib]{revtex4-2}

\usepackage{graphicx}
\usepackage{appendix}
\usepackage{amsmath}
\usepackage{amssymb}
\usepackage{xcolor}
\usepackage{mathrsfs}
\usepackage{hyperref}

\makeatletter
\renewcommand{\c@secnumdepth}{0}
\makeatother

\begin{document}

\preprint{APS/123-QED}

\title{Demonstration that Differential Length Changes of Optical Cavities\\ are a Sensitive Probe for Ultralight Dark Matter}

\author{Tejas Deshpande}
\author{Andra Ionescu}
\author{Nicholas Miller}
\author{Zhiyuan Wang}
\author{Gerald Gabrielse}
\author{Andrew A. Geraci}
\author{Tim Kovachy}
\affiliation{Center for Fundamental Physics, Department of Physics and Astronomy, Northwestern University, Evanston, Illinois 60208, USA}

\date{\today}

 \begin{abstract}
    Measurements of differential length  oscillations of Fabry-Perot cavities provide a sensitive and promising approach to searching for scalar ultralight dark matter (ULDM). The initial demonstration sets direct lower bounds that are one to two orders of magnitude lower for two model ULDM distributions -- a standard galactic halo and a relaxion star bound to Earth -- ranging over a decade of ULDM mass and Compton frequency. The demonstration suggests how a much higher sensitivity to a much larger ULDM mass range can be obtained.
\end{abstract}

\maketitle

Astronomical and cosmological evidence suggests that 85\% of the matter in the Universe is dark matter \cite{Bertone_2018_RMP}. It remains to be discovered whether the unidentified dark matter is described by the Standard Model (SM), or requires physics beyond the SM \cite{Safronova_2018_RMP}. Familiar axions \cite{Adams_2022_arXiv} and WIMPs \cite{Freese_2013_RMP} (weakly interacting massive particles) are proposed dark matter candidates that would also solve the so-called ``strong-CP'' \cite{Kim_2010_RMP} and ``hierarchy'' problems \cite{Dimopoulos_1981_PRD}.

\begin{figure}[htb!]
    \centering
    \includegraphics[width=\columnwidth]{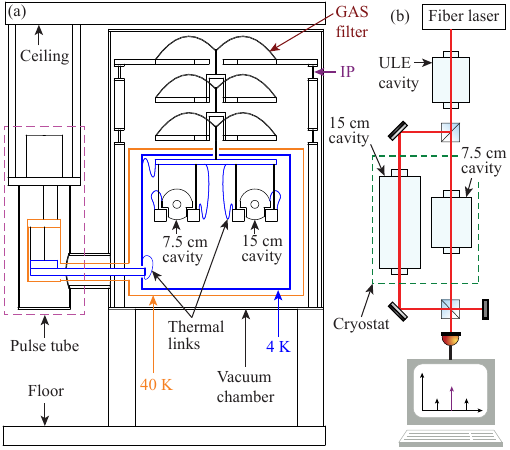}
    \caption{(a) Side view of the vibrationally-isolated cryostat.  (b) Schematic of the optical probe of the differential length variations of two cavities.}
    \label{fig:apparatus}
\end{figure}

This Letter reports a new approach to discovering or obtaining bounds on a well-motivated alternative -- scalar, parity-even, ultralight dark matter (ULDM) that behaves as a classical wave \cite{Ferreira_2021_AAR}. Fig.\,\ref{fig:apparatus} represents a demonstration apparatus to search for ULDM by comparing the lengths of two cavities with different mechanical responses to ULDM. This approach offers superior sensitivity to length changes at the ULDM Compton frequency, $f_{\phi} \equiv m_{\phi}c^{2}/(2\pi\hbar)$, in the 5\,kHz to 100\,kHz range, outperforming competing methods \cite{Savalle_2021_PRL,Tretiak_2022_PRL} by up to 2 orders of magnitude. Here, $m_{\phi}$, $\hbar$ and $c$ are the ULDM mass, Planck's constant and the speed of light respectively. Moreover, unlike other cavity-based schemes exploring $f_{\phi} \gtrsim 1\,{\rm kHz}$ \cite{Savalle_2021_PRL,Tretiak_2022_PRL}, the demonstration of a vibrationally isolated cryogenic setup opens pathways to significant sensitivity enhancement (up to 6 orders of magnitude) over a broader $f_{\phi}$ range (1\,kHz to 1\,MHz), thus uniquely allowing exploration of theoretically motivated parameter space. Furthermore, sensitivity to the local dark matter density produces strong and direct limits on boson stars \cite{Banerjee_2023_JHEP,Arakawa_2024_PRD} composed of relaxions -- a proposed solution to the hierarchy problem \cite{Banerjee_2020_CP} -- a contrast to the indirect limits from torsion balance equivalence principle (EP) tests \cite{Adelberger_2003_ARNPS,Antypas_2022_arXiv}.

ULDM is well-motivated, being consistent with the standard halo model (SHM) distribution of dark matter in a galaxy \cite{Antypas_2022_arXiv}, and naturally occurring in BSM (beyond the SM) theories. The topologically complex vacua of string theory, for example, produce many ULDM candidates. These include  moduli and dilatons  \cite{Dimopoulos_1996_PLB,Arkani-Hamed_2000_PRD,Burgess_2011_JHEP,Cicoli_2011_JHEP,Cicoli_2011_CQG,Taylor_1988_PLB,Damour_1994_NPB} that couple to matter by making SM parameters depend on moduli fields. The moduli can acquire mass through supersymmetry (SUSY) breaking at $\sim 0.1 \, {\rm meV}$ for TeV scale SUSY \cite{Dimopoulos_1996_PLB}, or have much lighter masses due to loop factors and small coefficients (e.g. for the electron Yukawa modulus). The electron mass and charge ($m_{\rm e}$ and ($-e$), and the fine structure constant ($\alpha$), vary in time ($t$) and space ($\mathbf{x}$) in proportion to the ULDM field $\phi(t,\mathbf{x})$,    
\begin{align}
\frac{\delta\alpha(t,\mathbf{x})}{\alpha}&=d_\alpha\frac{\sqrt{4\pi\hbar c}}{E_{{\scriptscriptstyle {\rm P}}}}\phi(t,\mathbf{x})\,,\nonumber\\
    \frac{\delta m_{{\rm e}}(t,\mathbf{x})}{m_{{\rm e}}}&=d_{m_{{\rm e}}}\frac{\sqrt{4\pi\hbar c}}{E_{{\scriptscriptstyle {\rm P}}}}\phi(t,\mathbf{x}) \, . \label{eq:DM_coupling}
\end{align}
$E_{{\scriptscriptstyle {\rm P}}} \equiv \sqrt{\hbar c^{5}/G} $ is the Planck energy and $G$ is Newton's gravitational constant. The dimensionless coupling constants are $d_{m_{{\rm e}}}$ and $d_{\alpha}$. The latter is often called $d_{{\rm e}}$ \cite{Arvanitaki_2015_PRD}, but $d_{\alpha}/2$ is the coupling to electron charge because $\alpha \propto e^2$.

The ULDM coherence length, $\lambda_{\phi} \equiv \hbar/(m_{\phi}v_{{\rm vir}})$ \cite{Derevianko_2018_PRA} is much larger than our tabletop apparatus for both ULDM models, so $\phi(t,\mathbf{x})\approx\phi(t)$ in Eq.\,(\ref{eq:DM_coupling}). The virial velocity $v_{{\rm vir}}$ is for ULDM in equilibrium, with mean kinetic energy equal to $(-1/2)$ multiplied by the mean gravitational potential energy. For the SHM, $v_{\rm vir} \approx 166 \, {\rm km}/{\rm s}$ (\cite{Freese_2013_RMP} and Appendix\,\ref{supp:data_analysis}) so $\lambda_{\phi}$ ranges from 8\,$\times 10^{2}$ to 172\,$\times 10^{2}$\,km for the $f_\phi$ between 5 and 100\,kHz on which we set new bounds. For a relaxion star gravitationally bound to the Earth, $v_{\rm vir} \approx 32 \, {\rm m}/{\rm s} \times (f_{\phi}/1 \, {\rm kHz})$ \cite{Banerjee_2020_CP} so $\lambda_{\phi}$ ranges from $5 \times 10^{4}$ to $110 \times 10^{4} \, {\rm km}$ for the $20$ to $90\,$kHz frequency range we consider.

Couplings to ULDM change $\alpha$ and $m_{\rm e}$, and hence the size of atoms and chemical bonds that go as the Bohr radius, $a_{{\scriptscriptstyle {\rm B}}} \equiv \hbar / (c \, m_{\rm e}  \, \alpha)$. The effective ULDM ``strain'' driving the length $L_{\rm cav}(t)$ of a rigid optical cavity is the fractional change in $a_{{\scriptscriptstyle {\rm B}}}$ \cite{Stadnik_2015_PRL,Stadnik_2016_PRA,Geraci_2019_PRL,Grote_2019_PRR}
\begin{equation}
    h_{{\scriptscriptstyle {\rm DM}}}(t) = -\frac{\delta m_{{\rm e}}(t)}{m_{{\rm e}}} - \frac{\delta\alpha(t)}{\alpha} \, . \label{eq:DM_strain}
\end{equation}
The resulting cavity strain, $h(t)=\delta L_{\rm cav}(t)/L_{\rm cav}$, will oscillate at $f_\phi$ for a continuous drive $h_{{\scriptscriptstyle {\rm DM}}}(t)$. Until a nonzero $h_{{\scriptscriptstyle {\rm DM}}}(t)$ is detected, limits on $\delta m_{{\rm e}}(t)/m_{{\rm e}}$ and $\delta \alpha(t)/\alpha$ are set in the usual way \cite{Antypas_2022_arXiv}, by assuming that $\delta \alpha(t)/\alpha = 0$ to place bounds on $\delta m_{{\rm e}}(t)/m_{{\rm e}}$ and vice versa. The former is reported. It applies for the latter.

\begin{figure}
    \centering
    \includegraphics[width=\columnwidth]{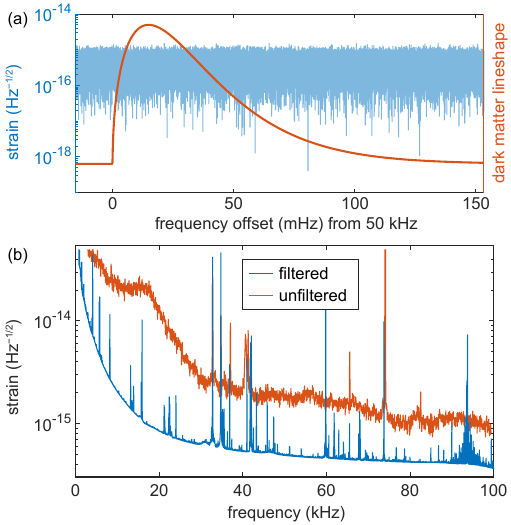}
    \caption{(a) Orange curve is a normalized analytic curve showing the expected ULDM lineshape in the SHM. Zoom in of the filtered strain ASD in (a) around 50\,kHz with $\approx 3\,\mu {\rm Hz}$ RBW. (b) The averaged ASD of the differential optical strain, $\widetilde{h}(f)$, with and without filtering through the ULE cavity.}
    \label{fig:beat}
\end{figure}

The ULDM coherence times $\tau_{{\rm c}} \equiv (c/v_{{\rm vir}})^{2}/(2\pi f_{\phi})$ \cite{Derevianko_2018_PRA,Centers_2021_ncomm} can range from short ($\tau_{\rm c} \ll T_{\rm m}$) to long ($\tau_{\rm c}\gg T_{\rm m}$) compared to our measurement time, $T_{\rm m} \approx 4\,\rm{days} \approx 4 \times 10^5\,\rm{s}.$ For the relaxion model, the coherence times for the range of $f_\phi$ we consider go from short ($\tau_{\rm c}=2 \times 10^{4}\,\rm{s}$) to long ($\tau_{\rm c}=2 \times 10^{6}\,\rm{s}$). For the SHM, the coherence time is always short since $\tau_c$ ranges only from $5\,\rm{s}$ to $10^2\,\rm{s}$ over the range of $f_\phi$ we consider.

For long coherence times, $\phi(t) \approx  \Phi_{0}\cos(2\pi f_{\phi} t + \theta)$ with a stochastic phase $\theta$, and amplitude $\Phi_{0} \equiv c\sqrt{2\rho_{{\scriptscriptstyle {\rm DM}}}}/(2\pi f_{\phi})$, that goes as the square root of the local dark matter density. For short coherence times, a sum over many Fourier components is required, in a frequency window around $f_{\phi}$ with a width that scales as $\tau_{\rm c}^{-1}$. All components go as $\sqrt{\rho_{{\scriptscriptstyle {\rm DM}}}}$. One result is the so-called ``ULDM lineshape'' \cite{Derevianko_2018_PRA}, an example of which is shown for the SHM in Fig.\,\ref{fig:beat}\,(a). All coherence times are appropriately accounted for in our Bayesian data analysis (\cite{Derevianko_2018_PRA,Centers_2021_ncomm} and Appendix \ref{supp:data_analysis}).

Fig.\,\ref{fig:apparatus} represents the apparatus used to search for ULDM Compton frequencies between 5 and 100\,kHz. Fig.\,\ref{fig:apparatus}\,(a) is an overview of the passive vibration isolation achieved via 5 stages: an inverted pendulum (IP), three geometric anti-spring (GAS) filters in series, and a 4-wire pendulum. A pulse tube refrigerator maintains the ``science cavities'' at 6\,K. The low temperature will be important for achieving higher sensitivities in the future.    

Both of the cryogenic optical cavities are made of single-crystal sapphire due to its high thermal conductivity, large Young's modulus ($\approx 400 \, {\rm GPa}$), and low intrinsic loss  at cryogenic temperatures \cite{Koroveshi_2023_PRD}. Their longitudinal axes are parallel to the crystallographic c-axis to maximize longitudinal stiffness. Optically contacted sapphire mirrors are coated with low-Brownian-noise crystalline GaAs/AlGaAs dielectric Bragg reflectors with Young's modulus and loss factor of $\approx 100 \, {\rm GPa}$ and $\approx 5 \times 10^{-6}$ respectively \cite{Cole_2013_nphot}.

Fig.\,\ref{fig:apparatus}\,(b) is an overview of the cavities and laser system. Amplified light from a fiber laser, with frequency $f_{{\rm opt}} \approx 193\,{\rm THz}$, is Pound-Drever-Hall locked to a $10 \, {\rm cm}$, 500,000-finesse ultralow expansion (ULE) cavity. The light is then split and frequency-shifted by $\approx \pm 131 \, {\rm MHz}$ (Fig.\,\ref{fig:apparatus_details}) to resonate with cavities of length $L_{\scriptscriptstyle{{\rm L}}} = 15\,{\rm cm}$ and $L_{\scriptscriptstyle{{\rm S}}} = 7.5\,{\rm cm}$ ($\approx 0.6\,{\rm mW}$ input per cavity). The frequency shift closes the gap in the cavities' optical resonances. The differential ``optical strain'' for fluctuations $\delta f_{{\rm beat}}(t)$ in their 262\,MHz beat note is
\begin{equation}
    h(t) \equiv \frac{\delta f_{{\rm beat}}(t)}{f_{{\rm opt}}}
    = A_{{\rm det}}(f_{\phi}) h_{{\scriptscriptstyle {\rm DM}}}(t) + h_{{\rm noise}}(t)
    \, . \label{eq:cavity_oscillation}
\end{equation}
This cavity response is the ``drive'' $h_{{\scriptscriptstyle {\rm DM}}}(t)$ times a detector response function $A_{{\rm det}}(f_{\phi})$. The added $h_{{\rm noise}}(t)$ from non-ULDM sources is critically reduced, from the orange to blue curves in Fig.\,\ref{fig:beat}\,(b), by filtering the laser through the ULE cavity. Laser frequency fluctuations offset from a cavity resonance by $f$ are suppressed as $1.5\,{\rm kHz}/ f$. Contributions to $h_{{\rm noise}}(t)$ from vibrations are greatly suppressed by the vibration isolation system.  

\begin{figure}
    \centering
    \includegraphics[width=\columnwidth]{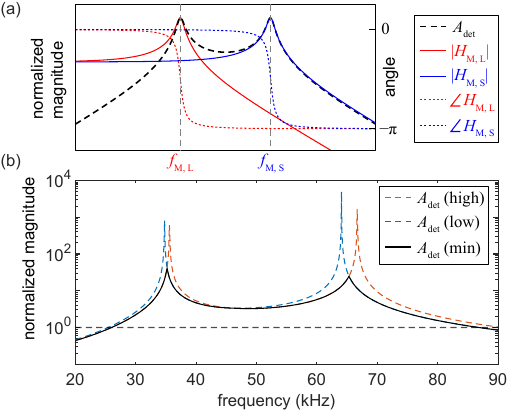}
    \caption{(a) Qualitative magnitude and phases of two  cavities' individual mechanical transfer functions (Eq.\,(\ref{eq:TF_mech})), $H_{{\scriptscriptstyle {\rm M,\,L/S}}}$ for the long/short cavity,  and the detector response function $A_{{\rm det}}$ (Eq.\,(\ref{eq:A_det})). (b) Dashed curves represent $A_{\rm det}$ for the lowest and highest data-driven estimates for the mechanical resonances of the two cavities with $Q_{{\scriptscriptstyle {\rm M}}} \sim 10^{4}$. The black curve conservatively takes their minimum to set ULDM limits.}
    \label{fig:transfer_functions}
\end{figure}

The ULDM $f_\phi$ primarily searched for are well above the optical poles of the long and short cavities. The detector response function of Eq.~(\ref{eq:cavity_oscillation}) then simplifies (Appendix\,\ref{supp:mechanical_resonances}) to the difference
\begin{equation}
    A_{{\rm det}}(f_\phi)=|H_{{\scriptscriptstyle {\rm M,\,L}}}(f_\phi)-H_{{\scriptscriptstyle {\rm M,\,S}}}(f_\phi)| \, , \label{eq:A_det}
\end{equation}
of two mechanical transfer functions  
\begin{equation}
    H_{{\scriptscriptstyle {\rm M}}}(f_{\phi}) = (1+Q_{{\scriptscriptstyle {\rm M}}}^{-2})^{1/2} \left(1-\left(\frac{f_{\phi}}{f_{{\scriptscriptstyle {\rm M}}}}\right)^{2}+{\rm i}Q_{{\scriptscriptstyle {\rm M}}}^{-1}\right)^{-1}.\label{eq:TF_mech}
\end{equation}
Each has its own resonant frequency $f_{{\scriptscriptstyle {\rm M}}}$ and quality factor $Q_{{\scriptscriptstyle {\rm M}}}$. The second subscript in Eq.~(\ref{eq:A_det}) designates the long (L) or short (S) cavity. Each sapphire cavity acts on a ULDM signal as a low pass filter, given that $Q_{{\scriptscriptstyle {\rm M}}}^{-1}$ is negligible away from $f_{{\scriptscriptstyle {\rm M}}}$. Above $f_{{\scriptscriptstyle {\rm M}}}$, the cavity does not follow the ULDM-induced oscillations \cite{Geraci_2019_PRL}.

\begin{figure}
    \centering
    \includegraphics[width=\columnwidth]{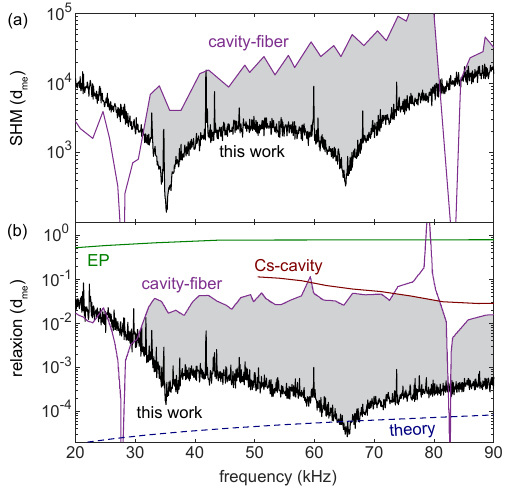}
    \caption{Bounds on $d_{m_{\rm e}}$ as a function of ULDM Compton frequency, for the SHM (a) and for a relaxion star bound to Earth (b), with limits from cavity-fiber \cite{Savalle_2021_PRL} and Cs-cavity \cite{Tretiak_2022_PRL} direct ULDM detectors, alongside indirect bounds from equivalence principle (EP) tests \cite{Antypas_2022_arXiv} and a theoretically motivated target \cite{Tsai_2023_nastro}.}
    \label{fig:d_me}
\end{figure}

\begin{figure}
    \centering
    \includegraphics[width=\columnwidth]{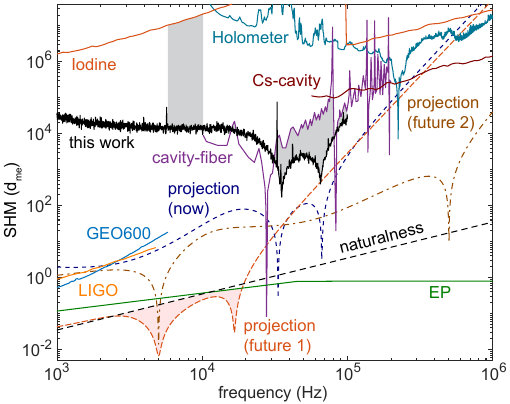}
    \caption{Bounds for the SHM for a much larger range of $d_{m_{\rm e}}$ and Compton frequencies compared to other bounds set by direct and indirect measurements, our sensitivity projections and theoretical limits. A zoom in of the dashed red box is shown in Fig.\,\ref{fig:d_me}\,(a). See main text for details.}
    \label{fig:d_me-LargeRange}
\end{figure}

We developed a data-driven model of Eq.~(\ref{eq:A_det}) by empirically estimating $f_{{\scriptscriptstyle {\rm M}}}$ and $Q_{{\scriptscriptstyle {\rm M}}}$ and fitting to finite-element analysis simulations. Fig.\,\ref{fig:transfer_functions} qualitatively shows magnitudes and phases of low pass transfer functions representing long ($f_{{\scriptscriptstyle {\rm M\,,L}}} \approx 34\,$kHz) and short ($f_{{\scriptscriptstyle {\rm M\,,S}}} \approx 64\,$kHz) cavities, whose difference determines $A_{\rm det}$. The short cavity approximately follows the ULDM drive, while the long cavity (largely unaffected by ULDM) is a length reference. The ULDM limits shown in Fig.\,\ref{fig:d_me} \& \ref{fig:d_me-LargeRange} account for the uncertainty in the model (Appendix\,\ref{supp:mechanical_resonances}).

Coherent ULDM sources drive the science cavities identically. A big advantage of the differential detection is that the strains of the two science cavities subtract within the absolute value in Eq.~(\ref{eq:A_det}), while uncorrelated noise sources $h_{{\rm noise}}(t)$ add in quadrature.

The data set for this work is a time series, $h(t)$, of measurements made every $4.63\,\mu\text{s}$ for 4\,days. Fourier transforming on a 16-core computer for 28\,hours produces the amplitude spectral density (ASD) $\widetilde{h}(f)$ in Fig.\,\ref{fig:beat}\,(b). All observed peaks are locally flat in that they are much wider than can be attributed to ULDM. The spectrum is well-described by white Gaussian noise in frequency windows approximately equal to the expected ULDM lineshape's spectral width. The data set is stationary in that the means, variances, and autocorrelation constants are independent of time, implying that neighboring frequency bins of $\widetilde{h}(f)$ are uncorrelated. The blue frequency spectrum in Fig.\,\ref{fig:beat}\,(a) is a raw ULE-cavity-filtered ASD of the differential optical strain, $\widetilde{h}(f)$, measured with a resolution bandwidth (RBW) $\approx 3 \, \mu {\rm Hz}$ around $f = 50\,{\rm kHz}$. The orange curve indicates the narrowness of the expected ULDM signal \cite{Derevianko_2018_PRA} for the SHM. The ASD of the differential optical strain is shown over a larger frequency range in the blue curve in Fig.\,\ref{fig:beat}\,(b), which is the average of $10^{5}$ spectra with a RBW of $\approx 0.3 \, {\rm Hz}$. To emphasize the greatly improved sensitivity achieved with ULE cavity filtering, the orange curve in  Fig.\,\ref{fig:beat}\,(b) (for a much smaller data set with 100\,Hz RBW and 25 averages) shows an ASD that is much higher due to unfiltered laser frequency noise.

The ASD of the large data set $\widetilde{h}(f)$ is analyzed to look for evidence of the two different ULDM models using Bayesian analysis \cite{Derevianko_2018_PRA,Centers_2021_ncomm}. In a frequency window as wide as the ULDM lineshape (e.g. Fig.\ \ref{fig:beat}\,(a)), we compute the likelihood that a ULDM signal $\widetilde{s}_{{\scriptscriptstyle {\rm DM}}}$ (Fourier transform of $A_{{\rm det}}h_{{\scriptscriptstyle {\rm DM}}}(t)$ for a given $d_{m_{{\rm e}}}$) gives rise to the measured data $\widetilde{h}(f)$. We use the likelihood function, $\mathcal{L}(\{\widetilde{h}\}|d_{m_{{\rm e}}})$, defined in Ref.\,\cite{Derevianko_2018_PRA,Centers_2021_ncomm}. The probability distribution function of the unknown parameter $d_{m_{\rm e}}$, known as the posterior, $\mathcal{P}(d_{m_{{\rm e}}}|\{\widetilde{h}\})$, is computed from the likelihood using Bayes theorem (Appendix \ref{supp:data_analysis}). Solving
\begin{equation}
    2\int_{0}^{d_{m_{{\rm e}}}^{95\%}}{\rm d}d_{m_{{\rm e}}}\,\mathcal{P}(d_{m_{{\rm e}}}|\{\widetilde{h}\}) = 0.95 \, . \label{eq:d_me_integral}
\end{equation}
gives $d_{m_{\rm e}}$ at a 95\% confidence level. The detector response function, $A_{{\rm det}}$, implicit in Eq.\,(\ref{eq:d_me_integral}) ensures the noise floor does not determine $d_{m_{{\rm e}}}$ in the parameter regime where the laser beam does not carry the dark matter signal.

In Fig.\,\ref{fig:d_me}\,(a), our increased sensitivity (gray area) and much smaller ULDM limits for the SHM (solid black curve) come from solving Eq.\,(\ref{eq:d_me_integral}). This is for a terrestrial detector moving through the galactic dark matter halo towards the Cygnus constellation, as the Solar system orbits the Milky Way, with a relative mean speed of $v_{\rm obs} \approx 230 \, {\rm km}/{\rm s}$ and a Gaussian velocity spread with $v_{\rm vir} \approx 166 \, {\rm km}/{\rm s}$ \cite{Freese_2013_RMP} (Appendix\,\ref{supp:data_analysis}). We look for spectral peaks in $\widetilde{h}(f)$ whose profile and 0.01 to 0.2\,Hz widths are set by the ULDM lineshape and inverse coherence time (Fig.\,\ref{fig:beat}\,(a)). For $T_{\rm m} \gg \tau_{{\rm c}}$, as is the case for the SHM, an approximate analytic solution to Eq.\,(\ref{eq:d_me_integral}) is $d_{m_{{\rm e}}} \approx 2.56 \, \widetilde{h}/(A_{{\rm det}}A_{0} \Phi_{0} (T_{\rm m}\,\tau_{{\rm c}})^{1/4})$, where $A_{0}\equiv \sqrt{4\pi\hbar c}/E_{{\scriptscriptstyle {\rm P}}}$ \cite{Derevianko_2018_PRA}. The $f_{\phi}^{-1}$ scaling of $\Phi_{0}$ and $\tau_{{\rm c}}$ and accounting for the optical poles in $A_{{\rm det}}$ (Appendix\,\ref{supp:mechanical_resonances}) compensates for the increase in $\widetilde{h}$ (Fig.\,\ref{fig:beat}\,(b)) at lower frequencies.

In Fig.\,\ref{fig:d_me}\,(b), our increased sensitivity (gray area) and much smaller ULDM limits (solid black curve) for the model of a relaxion star gravitationally bound to Earth are solutions to Eq.\,(\ref{eq:d_me_integral}). The density of this relaxion halo at our detector is a factor of $10^{11}$ to $10^{13}$ greater than its SHM counterpart \cite{Banerjee_2020_CP}, thereby enhancing the limits from our direct ULDM detector by approximately 6 orders of magnitude as compared to the SHM case. For ULDM Compton frequencies less than $34\,{\rm kHz}$, for which $\tau_{\rm c} > T_{\rm m}$, the $d_{m_{{\rm e}}}$ bound is higher due to the stochastic fluctuation of the ULDM amplitude, $\Phi_{0}$, that is included in the Bayesian analysis \cite{Centers_2021_ncomm}. In the 64 to 66\,kHz range, the bound exceeds a target corresponding to the boundary of the region of parameter space for which Higgs-relaxion mixing can occur \cite{Tsai_2023_nastro}. 

For both models, between 20 to 90\,kHz there is no evidence of ULDM above the noise floor of $\widetilde{h}(f)$. Peaks visible in Fig.\,\ref{fig:beat}\,(b) are much broader than expected for a ULDM signal. They likely originate from technical noise sources and thus are part of the noise floor that provides exclusion limits on $d_{m_{\rm e}}$. The previous best limits for both models come from comparing the lengths of a single cavity at different times using an optical fiber delay line (`cavity-fiber' curve in Fig.\,\ref{fig:d_me}) \cite{Savalle_2021_PRL}.

Fig.\,\ref{fig:d_me-LargeRange} displays our SHM result for a much broader range of coupling constants and Compton frequencies to relate the new bounds to other measurement bounds and theoretical constraints.  In addition to those shown in Fig.\,\ref{fig:d_me}\,(a) and (b), these include direct ULDM limits from gravitational wave (GW) detectors \cite{Vermeulen_2021_nature,Gottel_2024_PRL} and the Holometer \cite{Aiello_2022_PRL}, along with indirect limits from EP tests \cite{Antypas_2022_arXiv,Schlamminger_2008_PRL,Wagner_2012_CQG}. Radiative corrections to the ULDM mass provide a very model-dependent upper limit on $d_{m_{\rm e}}$, with the parameter space above the dashed line from \cite{Antypas_2022_arXiv} labeled ``naturalness'' excluded without substantial fine-tuning \cite{Geraci_2019_PRL}. In addition, we set lower limits in the 6 to 10\,kHz range for the SHM, where the previous strongest limits came from molecular Iodine spectroscopy \cite{Oswald_2022_PRL}.

Furthermore, Fig.\,\ref{fig:d_me-LargeRange} shows theoretical $d_{m_{\rm e}}$ projections limited by the cavities' thermal and shot noise for the current (``projection (now)'') and proposed future apparatuses (Appendix\,\ref{supp:noise_projections}). The latter is shown for two configurations $L_{\scriptscriptstyle{{\rm L}}} = 1\,{\rm m}, \, L_{\scriptscriptstyle{{\rm S}}} = 30\,{\rm cm}$ (``projection (future 1)'') and $L_{\scriptscriptstyle{{\rm L}}} = 1\,{\rm m}, \, L_{\scriptscriptstyle{{\rm S}}} = 1\,{\rm cm}$ (``projection (future 2)''). Projections assume temperature, quality factor ($\approx 10^{4}$), and finesse similar to our current apparatus (Appendix\,\ref{supp:apparatus}). However, measurement time ($T_{\rm m} = 10^{7} \, {\rm s}$) and input laser power (50\,mW per cavity) were increased. Cavity lengths affect the explored parameter space; the first configuration prioritizes sensitivity over bandwidth, exploring below the naturalness line. We attribute the discrepancy in the measured and theoretical sensitivities to residual laser frequency/amplitude noise, which we plan to mitigate with additional optical filtering in future apparatuses.


Our projected sensitivities highlight the advantage of our method over other cavity-based ULDM detection schemes in Fig.\,\ref{fig:d_me-LargeRange}. Schemes involving comparison of cavities to atoms \cite{Tretiak_2022_PRL} or molecules \cite{Oswald_2022_PRL} are limited by factors like lower atom flux and broad atomic transition linewidths. Cavities benefit from a much higher photon flux, lower cavity linewidth and low thermal Brownian noise. While the cavity in Ref.\,\cite{Savalle_2021_PRL} shares these benefits, their self-referential nature results in loss of sensitivity to ULDM above the cavity's mechanical resonance. Moreover, fiber delay lines suffer from acoustic and thermal noise of the long fiber at low frequencies. Dual-cavity optical interferometers like GW detectors do not improve ULDM sensitivity with extended Fabry-Perot cavities. Since only beam splitter oscillations are monitored, their sensitivity is fundamentally comparable to their tabletop counterparts \cite{Grote_2019_PRR}.


In conclusion, a demonstration of a fresh approach to broadband direct ULDM detection sets new limits on the coupling of ultralight dark matter (ULDM) and ordinary matter.  The new bounds are up to two orders of magnitude lower for both the standard halo model (SHM) and for a relaxion star bound to Earth. Routes to sensitivity increases to exceed the naturalness bound and a much broader range of ULDM masses now seem feasible.      

\emph{Acknowledgments:} The Templeton Foundation funded the apparatus construction and the Packard Foundation funded T.\,K.; B.\,Odom contributed to early stages of this work and useful suggestions came from C.\,Schreibeis, A.\,Derevianko, G.\,Cole, S.\,Catano, R.\,DeSalvo, K.\,Arai, Y.\,Michimura, G.\,Venugopalan, and J.\,Bourhill. The custom dewar, inverted pendulum, and geometric anti-spring system were built by Four Nine Design to our specifications.

\clearpage

\appendix


\begin{figure}
    \centering
    \includegraphics[width=\columnwidth]{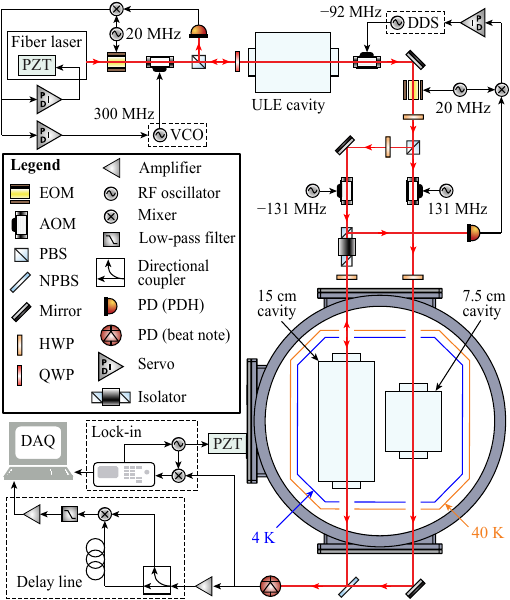}
    \caption{Detailed optical layout of the cavities' beat note and mechanical resonance measurement apparatus.}
    \label{fig:apparatus_details}
\end{figure}


\section{Isolated Sapphire Cavities at 6\,K}
\label{supp:apparatus}

Four stages of vibration isolation, each with a mechanical pole $\approx 0.5 \,{\rm Hz}$, are within the dewar vacuum but operate at 300\,K (Fig.\,\ref{fig:apparatus}\,(a)). An inverted pendulum (IP), composed of 3 flexures, provides horizontal isolation. The IP supports a stack of 3 geometric anti-spring (GAS) filters that provide vertical isolation after the mass that each supports is tuned within a few grams to produce the mentioned pole frequency. Each GAS is 3 blade springs that meet at a point, from which the stage below is supported. 

The fifth stage is a cryogenic 4-wire pendulum with a $\approx 1.5 \,{\rm Hz}$ mechanical pole. The pendulum's Molybdenum wires are $250\,\mu {\rm m}$ in diameter and 15\,cm long. Each pendulum supports an aluminum plate that, in turn, supports one optical cavity. The optical cavity spacers are supported by the aluminum platforms at 4 points that finite element calculations indicated would minimize vibration transmission \cite{Nazarova_2006_APB,Andra_thesis}.  

Heat is extracted from the cavities using high conductivity aluminum links (0.1\,mm by 1\,cm by 10\,cm). They are very pure and annealed to maximize the heat transfer, and to make them as flexible as possible to minimize vibration transmission. They are bonded to the sapphire using silver paint and clamped to the cavity platforms. Four-wire silicon diode sensors attached to the cavities with silver paint typically measure 6\,K.

\begin{figure}[htb!]
    \centering
    \includegraphics[width=1\columnwidth]{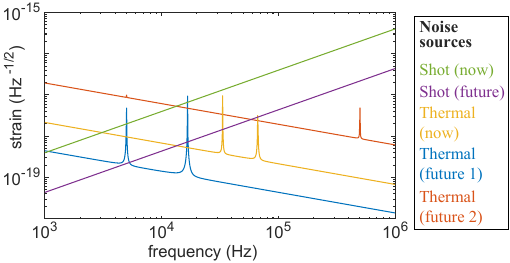}
    \caption{Theoretical contributions of the thermal and shot noise for the sensitivity projections shown in Fig.\,\ref{fig:d_me-LargeRange}.}
    \label{fig:therm_shot_noise}
\end{figure}

\section{Optics and Readout of Cavity Length}
\label{supp:optics}

Fig.\,\ref{fig:apparatus_details} gives more optics details for the PDH locking to the ULE cavity, for the frequency shifting to send nearly resonant light through each sapphire cavity, and for slowly compensating the very small frequency drift of these cavities. The fiber laser's internal piezoelectric transducer (PZT) is used for slow feedback (1\,kHz) in its PDH lock to the ULE cavity. The fiber amplifier is required because the 40\,mW power of the fiber laser was otherwise reduced by the 20\,dB attenuation of the narrow band ULE cavity setup, enough to cause the detected signal after the science cavities to be compromised by detector noise. Electro-optic modulators (EOMs) and acousto-optic modulators (AOMs) at the frequencies indicated in the figure are used in usual ways for PDH locking and shifting the laser frequency, along with half- and quarter-wave plates (HWPs and QWPs) and polarizing beam splitters (PBSs) for polarization control. For brevity, the AOMs' double-pass and mode-matching lenses are not shown. The AOMs are driven by analog voltage-controlled oscillators (VCOs) and direct digital synthesizers (DDSs).

The laser beams after the sapphire cavities are recombined on a 50:50 non-polarizing beam splitter (NPBS) and detected with a 5\,GHz beat note photodetector (PD) from Thorlabs (DET08C). The 262\,MHz carrier is canceled by mixing with the same output after a delay line \cite{HP_delay_line} and the resulting time series (i.e. $\delta f_{{\rm beat}}(t)$ from Eq.\,(\ref{eq:cavity_oscillation})) is stored on a data acquisition (DAQ) card. A Fourier transform of the time series from this ``self-homodyne detection'' corresponds to the optical strain spectrum (Fig.\,\ref{fig:beat}) as a function of the offset frequency from the carrier, insensitive to slow drifts of the carrier frequency. The discriminator is calibrated using the signal amplitude when an optical sideband is added with one of the AOMs (see \cite{HP_delay_line} for calibration procedure). A summary of important apparatus parameters is shown in Table\,\ref{tab:values}.


\begin{table}[h]
    \begin{centering}
    \begin{tabular}{|l|l|}
        \hline
        \textbf{Description (notation)} & \textbf{Values} \tabularnewline
        \hline
        Cavity lengths ($L_{{\scriptscriptstyle {\rm L/S}}}$) & 15\,cm / 7.5\,cm \tabularnewline
        \hline
        Mechanical resonances ($f_{{\scriptscriptstyle {\rm M,\,L/S}}}$) & $34\,{\rm kHz}$ / $64\,{\rm kHz}$ \tabularnewline
        \hline
        Mechanical quality factor ($Q_{{\scriptscriptstyle {\rm M}}}$) & $10^4$ \tabularnewline
        \hline
        Optical poles ($f_{{\rm p,}\,{\scriptscriptstyle {\rm L/S}}}$) & $4.7\,{\rm kHz}$ / $7.4\,{\rm kHz}$ \tabularnewline
        \hline
        Optical frequency ($f_{\rm opt}$) & $193 \, {\rm THz}$ \tabularnewline
        \hline
        Input power per cavity & 0.6\ mW \tabularnewline
        \hline
        Cavity power transmission efficiency & 20\% \tabularnewline
        \hline
        Cavity finesse & 120,000 ($\pm 12\%$) \tabularnewline
        \hline
    \end{tabular}
    \end{centering}
    \caption{Summary of important apparatus parameter with nominal values.}
    \label{tab:values}
\end{table}

\section{Theoretical limit estimation}
\label{supp:noise_projections}

The sensitivity projections in Fig.\,\ref{fig:d_me-LargeRange} were computed using the theoretical estimates of the cavities' thermal and shot noise (Fig.\,\ref{fig:therm_shot_noise}). We assume our vibration isolation upgrades will be comparable to GW detectors where such noise is subdominant above $100 \, {\rm Hz}$ \cite{Sekiguchi_thesis}. Therefore, we conservatively ignore vibrational noise for $f_{\phi} > 1\,{\rm kHz}$. Thermal noise is modeled by the formulas in \cite{Andra_thesis}. The shot noise strain, $\widetilde{h}_{\rm shot}(f_{\phi}) = f_{\phi}(2\pi\hbar(P_{{\scriptscriptstyle {\rm L}}}^{-1} + P_{{\scriptscriptstyle {\rm S}}}^{-1})/(\eta f_{\rm opt}))^{1/2}$, is determined by the powers ($P_{{\scriptscriptstyle {\rm L}}}$ and $P_{{\scriptscriptstyle {\rm S}}}$) from the long and short cavities incident on the photodetector, with quantum efficiency $\eta \approx 0.7$. $P_{{\scriptscriptstyle {\rm L}}}$ and $P_{{\scriptscriptstyle {\rm S}}}$ are balanced using a HWP before the PBS (Fig.\,\ref{fig:apparatus_details}). A second photodetector at the beam splitter's other port can further improve the shot noise limit. Future increases in input power to 50\,mW per cavity ($P_{{\scriptscriptstyle {\rm L/S}}} \approx 5\,{\rm mW}$ with 20\% transmission) will not significantly increase thermal noise. Conservatively, assuming a cooling power no lower than now ($\approx (8 \, {\rm mK}/{\rm mW})^{-1}$), we estimate a $\approx 0.4 \, {\rm K}$ temperature increase.

\section{Detector response}
\label{supp:mechanical_resonances}

A more general detector response function,
\begin{equation}
    A_{{\rm det}}=|H_{{\scriptscriptstyle {\rm M,\,L}}}H_{{\scriptscriptstyle {\rm OM,\,L}}}-H_{{\scriptscriptstyle {\rm M,\,S}}}H_{{\scriptscriptstyle {\rm OM,\,S}}}| \, . \label{eq:A_det_full}
\end{equation}
is used in our numerical analysis rather than Eq.\,(\ref{eq:A_det}). The optomechanical transfer function is
\begin{equation}
    H_{{\scriptscriptstyle {\rm OM}}}(f_{\phi})=-\frac{{\rm i}f_{\phi}}{{\rm i}f_{\phi}+f_{{\rm p}}} \, , \label{eq:TF_optomech}
\end{equation}
for a cavity pole frequency $f_{{\rm p}}$ \cite{Izumi_2012_KAGRA,Andra_thesis}. For $f_{\phi} \gg f_{{\rm p}}$, since $H_{{\scriptscriptstyle {\rm OM}}} \approx 1$, Eq.\,(\ref{eq:A_det}) is a good approximation that is useful for an intuitive understanding.

There is uncertainty in our determination of the optical pole frequencies, and also of the mechanical resonance frequencies and quality factors of the two science cavities. The optical poles ($f_{{\rm p,}\,{\scriptscriptstyle {\rm L/S}}}$) are estimated by fitting to the cavities' measured optical transfer function. The mechanical resonances ($f_{{\scriptscriptstyle {\rm M,\,L/S}}}$) are empirically estimated by observing the enhancement of peaks in the ASD resulting from a PZT drive (Fig.\,\ref{fig:apparatus_details}). Mechanical quality factors ($Q_{{\scriptscriptstyle {\rm M}}}$) are estimated by measuring the full width at half maximum of the resonance peaks (Fig.\,\ref{fig:transfer_functions}\,(b)). Measurements of $f_{{\scriptscriptstyle {\rm M}}}$ and $Q_{{\scriptscriptstyle {\rm M}}}$ serve as calibration for finite element analysis simulating a volumetric ULDM drive which agrees with the form of Eq.\,(\ref{eq:TF_mech}).

\section{Data Acquisition and Processing}
\label{supp:data_analysis}

The DAQ card rated to 216\,kS/s recorded $N = 7.48 \times 10^{10}$ samples with 5 significant figures at a sampling interval $\Delta t = 4.63 \, \mu{\rm s}$. The ASD, $\widetilde{h}(f_{p})$, is a function of the discrete frequency $f_{p} \equiv p/(N\Delta t)$ for $p \in [\![0,N/2]\!]$.

For $N \Delta t \gg \tau_{\rm c}$ the power spectral density of the optical strain (i.e. our signal) is
\begin{equation}
    \widetilde{S}_{p}=d_{m_{{\rm e}}}^{2}A_{{\rm det}}^{2}\langle|\widetilde{\phi}_{p}|^{2}\rangle \, . \label{eq:DM_PSD}
\end{equation}
where $\widetilde{\phi}_{p}$ is the discrete Fourier transform of the ULDM field and $\langle \ldots \rangle$ denote a statistical average. Note that all of the above quantities are implicit functions of the ULDM Compton frequency $f_{\phi}$. Moreover, $\langle|\widetilde{\phi}_{p}|^{2}\rangle$ is defined in terms of the ULDM lineshape function $F_{p}$ as \cite{Derevianko_2018_PRA}
\begin{equation}
    \langle|\widetilde{\phi}_{p}|^{2}\rangle\equiv\frac{\pi N}{\Delta t}\Phi_{0}^{2}F_{p} \, , \label{eq:DM_field_correlation}
\end{equation}
where $\Phi_{0} \equiv c\sqrt{2\rho_{{\scriptscriptstyle {\rm DM}}}}/(2\pi f_{\phi})$. Assuming a 3D velocity distribution of the ULDM to be $f_{{\scriptscriptstyle {\rm DM}}}(\mathbf{v})=(2\pi v_{{\rm vir}}^{2})^{-3/2}\exp(-(\mathbf{v}-\mathbf{v}_{{\rm obs}})^{2}/(2v_{{\rm vir}}^{2}))$, and ignoring annual modulation, we get
\begin{align}
    F_{p} &\equiv \frac{\tau_{{\rm c}}}{\sqrt{2\pi}\,\eta}{\rm e}^{-\eta^{2}-2\pi(f_{p}-f_{\phi}^{\prime})\tau_{{\rm c}}} \nonumber \\
    &\times \sinh\left(\eta\sqrt{\eta^{2}+4\pi(f_{p}-f_{\phi}^{\prime})\tau_{{\rm c}}}\right) \, , \label{eq:DM_lineshape}
\end{align}
where $\tau_{{\rm c}}\equiv\hbar/(m_{\phi}v_{{\rm vir}}^{2})$ is the coherence time, $f_{\phi}^{\prime} = f_{\phi} + m_{\phi} v_{\rm obs}^{2}/(4\pi\hbar)$, and $\eta \equiv v_{\rm obs}/v_{\rm vir}$ ($v_{{\rm obs}}\equiv|\mathbf{v}_{{\rm obs}}|$). For the SHM ($v_{\rm obs} \approx 230\,{\rm km}/{\rm s}$ and $v_{\rm vir} \approx 166 \, {\rm km}/{\rm s}$), $\eta \approx 1$ \cite{Freese_2013_RMP}. For the relaxion star bound to Earth (i.e. $v_{\rm obs} \approx 0$), $\eta \approx 0$. For the latter, $\rho_{{\scriptscriptstyle {\rm DM}}}$ and $v_{\rm vir}$ become dependent on $f_{\phi}$ \cite{Banerjee_2020_CP}.

Because $f_{\phi}$ is a large number of ULDM linewidths ($\sim \tau_{\rm c}^{-1}$) away from zero ($p=0$) and from the Nyquist ($p=N/2$) frequency, we can evaluate the likelihood function \cite{Derevianko_2018_PRA}
\begin{equation}
    \mathcal{L}(\{\widetilde{d}_{p}\}|d_{m_{{\rm e}}})\equiv\prod_{p=1}^{N/2-1}\frac{1}{\pi\widetilde{\Sigma}_{p}}\exp\left[-\frac{|\widetilde{d}_{p}|^{2}}{\widetilde{\Sigma}_{p}}\right] \, , \label{eq:likelihood_marginalized}
\end{equation}
where $\widetilde{d}_{p}\equiv(N/\Delta t)^{1/2} \, \widetilde{h}(f_{p})$ and $\widetilde{\Sigma}_{p}\equiv\widetilde{\rho}_{p}+\widetilde{S}_{p}$. The detector's noise model $\widetilde{\rho}_{p}$ is the averaged $|\widetilde{d}_{p}|^{2}$.

The posterior can be computed from Eq.\,(\ref{eq:likelihood_marginalized}) using Bayes theorem
\begin{equation}
    \mathcal{P}(d_{m_{{\rm e}}}|\{\widetilde{d}_{p}\})=\frac{p(d_{m_{{\rm e}}})}{p(\{\widetilde{d}_{p}\})}\mathcal{L}(\{\widetilde{d}_{p}\}|d_{m_{{\rm e}}}) \, . \label{eq:Bayes_theorem}
\end{equation}
We assume an uninformed prior, $p(d_{m_{{\rm e}}})$, and $p(\{\widetilde{d}_{p}\})$ is a normalization constant.

\end{document}